# Elastic and electronic properties of PbO-type FeSe$_{1-x}$Te$_x$ (x = 0 – 1.0): A first-principles study

S. Chandra, A.K.M.A. Islam[1]

*Department of Physics, Rajshahi University, Rajshahi, Bangladesh*

**ABSTRACT**

The effect of doping on electronic and other related properties of PbO-type FeSe$_{1-x}$Te$_x$ has been investigated theoretically using density functional method. The elastic properties for mono- and poly-crystalline FeSe$_{1-x}$Te$_x$ system are predicted for the first time and the results discussed. Analysis of doping dependent band characteristics in conjunction with previous studies reveal that favorable nesting of Fermi surface indicates a possible basis for understanding why $T_c$ in FeSe$_{1-x}$Te$_x$ is maximum for $x \sim 0.5$.

## 1. Introduction

The discovery of LaOFeAs as a high temperature superconductor [1] has initiated much research on iron-based layered superconductors [2-11]. At least three types of crystal structures have been identified for these superconductors as: (i) the ZrCuSiAs-type layered tetragonal structure for LaOFeAs [1, 2], (ii) relatively simpler ThCr$_2$Si$_2$-type layered tetragonal structure for BaFe$_2$As$_2$ [3, 4], and (iii) the simplest PbO-type layered tetragonal structure for FeSe [5]. The last phase, known as α-FeSe, has been reported to be a superconductor with transition temperature, $T_c \sim 8$ K (with some Se deficiency) [5, 6]. A large enhancement of $T_c = 27$ K was observed at pressure $P \sim 1.48$ GPa [6]. A more recent experiment [7] shows that the application of hydrostatic pressure first rapidly increases $T_c$ which attains a broad maximum of 37 K at ~7 GPa before decreasing to 6 K upon further compression to ~14 GPa. This is one of the highest $T_c$ ever reported for a binary solid. Since the discovery the tetragonal iron selenide has attracted more and more attention, motivated in large part by the peculiar electronic, optical, and magnetic properties, and also as a promising functional material due to its relevance with (Fe-atom) spintronics [12].

The newly discovered superconductor FeSe has been subjected to theoretical investigation by Subedi *et al.* [8]. On the other hand the compounds FeSe$_{1-x}$Te$_x$ (*x* = 0-1), where Te substitution has an effect on superconductivity, has been investigated experimentally [14-16]. It was found that $T_c$ increases with Te doping, reaching a maximum $T_c \sim 15$ K at about 50-70% substitution, and then decreases with more Te doping. This compositional dependence of $T_c$, decreasing for both underdoped and overdoped materials has also been observed in the cuprates (see [17]).

On the theoretical side electronic properties (e.g. band structure) have only been done for FeSe (*x* = 0) and FeTe (*x* = 1.0) [8], but not for intermediate *x* concentrations. The replacement of Se by Te introduces a chemical pressure into the structure of the Fe(Se$_{1–x}$Te$_x$) system that may have similar effects on the electron clouds to those due to a mechanical pressure. Very recently during the course of the work we came across a paper by Singh [18] in which the effects of excess Fe, Se deficiency and substitutions of S and Te on the Se sub-lattice and Co, Ni and Cu on the Fe sub-lattice in FeSe has been studied using the coherent-potential approximation. A systematic theoretical examination of the effect of Te doping on electronic and other related properties in FeSe$_{1-x}$Te$_x$ system at different compositional levels, including for $x$ value for which $T_c$ is maximum, may yield new information. Further we know that the elastic constants are related to mechanical and dynamical properties from which one can obtain valuable information of material such as chemical bonding, stability and stiffness, specific heat, thermal expansion, Debye temperature and so on. The estimation of Debye temperature is useful in evaluating the electron-phonon coupling constant $\lambda$, which is proportional to the mean sound velocity $v_m$. Thus the purpose of our work would also be to study the elastic properties of the FeSe$_{1-x}$Te$_x$ system for which there are no elastic constants available. The effect of Te concentration on the compounds on the strength of coupling between planes will also be discussed.

---

[1] *Corresponding author*: azi46@ru.ac.bd

## 2. Methods of calculation

Our calculations are performed using the plane-wave pseudopotential method within the fame work of the density functional theory [19] implemented in CASTEP code [20]. The exchange correlation term is considered by the local density approximation (LDA) parameterized by Ceperly-Alder [21]. We have used a 10×10×8 Monkhorst grid to sample the Brillouin zone. We set the plane-wave cutoff energy to be 600 eV and the convergence of the force on each atom to be 0.01 eV/Å. The optimization of the lattice constants and the atom coordinates is made by minimization of the total energy.

## 3. Results and discussions

### 3.1. *Geometrical optimization*

α-FeSe crystallizes at room temperature with the tetragonal PbO-type structure with 4 atoms/cell (space group *P4/nmm*) as shown in Fig. 1. The structure consists of stacks of edge-sharing FeSe$_4$ tetrahedra, the FeSe packing pattern is essentially identical to that of the FeAs layers in the iron oxyarsenides. Table 1 presents the calculated lattice constants, internal coordinate, bond lengths and bond angles for $FeSe_{1-x}Te_x$ ($x = 0 - 1$) at zero temperature. The available experimental data at room temperature and other theoretical data are also included in the table. It is found that all the optimized parameters predicted by the DFT method are in reasonable agreement with the experimental data [14, 22].

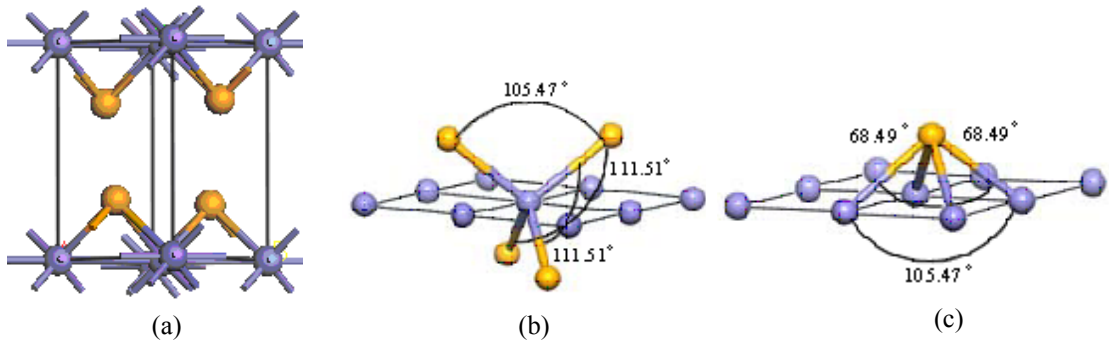

(a)          (b)          (c)

Fig.1. (a) α-FeSe structure. Fe and Se ions are depicted as blue and yellow spheres, respectively. Geometry of the FeSe$_4$ tetrahedra and (c) the SeFe$_4$ pyramids with the three distinct Se-Fe-Se and Fe-Se-Fe bond angles are indicated.

Table1. Optimized lattice constants, internal coordinate, bond lengths and bond angles for $FeSe_{1-x}Te_x$ ($x = 0 - 1$).

| Compound | $a$ (Å) | $c$ (Å) | $z$ | Bond length (Å) | | Bond angle (°) | |
|---|---|---|---|---|---|---|---|
| | | | | Fe–Fe:4 | (Se/Te)-Fe:4 | (Se/Te)-Fe-(Se/Te) | Fe-(Se/Te)-Fe |
| [a]FeSe | 3.7752 | 5.5268 | 0.26 | 2.670 | 2.373 | 105.43 | 68.52 |
| [b]FeSe | 3.775* | 5.512* | - | - | - | - | - |
| [c]FeSe | 3.656 | 5.375 | 0.259 | 2.585 | 2.297 | 105.47 | 68.49 |
| [d]FeSe | 3.58 | 5.31 | 0.26 | - | - | - | - |
| [a]$FeSe_{0.75}Te_{0.25}$ | 3.7872 | 5.6492 | 0.26 | 2.678 | 2.397 | 112.12 | 67.93 |
| [c]$FeSe_{0.75}Te_{0.25}$ | 3.6816 | 5.5658 | 0.263 | 2.603 | 2.352 | 112.79 | 67.21 |
| [a]$FeSe_{0.50}Te_{0.50}$ | 3.7913 | 5.9784 | 0.26 | 2.681 | 2.451 | 113.73 | 66.33 |
| [b]$FeSe_{0.50}Te_{0.50}$ | 3.7924* | 5.946* | - | - | - | - | - |
| [c]$FeSe_{0.50}Te_{0.50}$ | 3.7115 | 5.7118 | 0.265 | 2.624 | 2.396 | 113.58 | 66.42 |
| [a]$FeSe_{0.25}Te_{0.75}$ | 3.8129 | 6.1500 | 0.26 | 2.696 | 2.488 | 114.43 | 65.63 |
| [c]$FeSe_{0.25}Te_{0.75}$ | 3.7397 | 5.8356 | 0.266 | 2.644 | 2.432 | 114.13 | 65.87 |
| [a]FeTe | 3.8266 | 6.2935 | 0.26 | 2.706 | 2.559 | 115.02 | 65.02 |
| [b]FeTe | 3.8275* | 6.2682* | - | - | - | - | - |
| [c]FeTe | 3.7634 | 5.9322 | 0.268 | 2.661 | 2.462 | 114.58 | 65.42 |

[a]Expt. [22], [b]Expt. [14] ; *From graph [14] ; [c]This; [d][23] using VASP.

The $c$- and $a$-axis lengths against Te concentration are shown in Fig. 2 in comparison with the lattice expansion obtained from X-ray refinement at room temperature [22]. We find an average of 4.0%, and 2.3% differences between the calculated and experimental $c$ and $a$ axis-lengths, respectively as we go from $x = 0$ to $x = 1.0$. The larger axis-lengths for different $x$-values obtained in the X-ray data are due to the fact that the data correspond to room temperature and those in the theory correspond to zero temperature. The calculation shows that $c$-axis expands by ~3.4% as we increase $x$ from zero to 0.25. The expansion becomes more than 10.4% as $x$ approaches 1.0, which may be compared with 12% found in experiment [22]. The corresponding expansion in $a$ for $x = 0$ to $x = 1.0$ is only ~2.9%. Thus we see that Te doping increases the spacing between neighbouring Fe-occupied planes. The larger ionic radius of Te, as compared to that of Se, the Fe-Fe and Fe-(Se,Te) bond lengths also increase by ~3% and ~7%, respectively (see Table 1). Thus higher concentration of Te induces weaker coupling between each plane and results in more layer-like characteristics [16]. The geometry of the $FeSe_4$ tetrahedra and the $SeFe_4$ pyramids are seen to change progressively (Fig. 1 and Table 1) as the substitution of Se by Te increases.

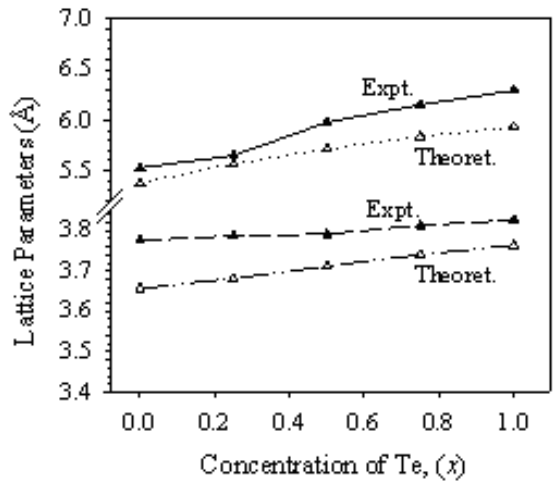

Fig. 2. Lattice parameters $a$ (lower two curves) and $c$ (upper two curves) as a function of concentration of $FeSe_{1-x}Te_x$. The experimental data are from [22].

## 3.2 Elastic properties

### 3.2.1 Elastic parameters of mono- and poly-crystalline $FeSe_{1-x}Te_x$

In the present work, we performed systematic first-principles calculations of the elastic parameters of $FeSe_{1-x}Te_x$ single crystals, such as the elastic constants $C_{ij}$, the bulk moduli $B$ and the shear moduli $G$. These are widely used for describing the elastic behaviour of materials. First $C_{ij}$'s are evaluated by calculating the stress tensors on different deformations applied to the equilibrium lattice of the tetragonal unit cell. Then the dependence between the resulting energy change and the deformation are determined, and the constants are evaluated in a standard way.

The calculated six independent elastic coefficients ($C_{11}$, $C_{12}$, $C_{13}$, $C_{33}$, $C_{44}$, and $C_{66}$) of $FeSe_{1-x}Te_x$ for each doping level are found to be positive and satisfy the known Born criterion for a mechanically stable system: $C_{11} > 0$, $C_{33} > 0$, $C_{44} > 0$, $C_{66} > 0$, $(C_{11} - C_{12}) > 0$, $(C_{11} + C_{33} - 2C_{13}) > 0$, and $\{2(C_{11} + C_{12}) + C_{33} + 4C_{13}\} > 0$.

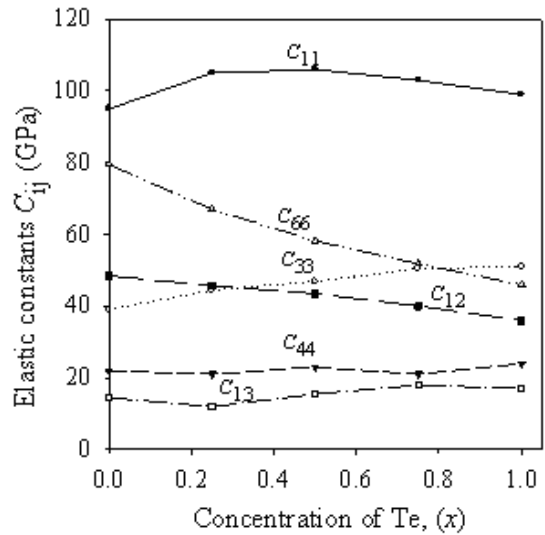

Fig. 3. Variation of the elastic constants $C_{ij}$ with Te concentration ($x$).

Fig 3 shows the variation of elastic constants with Te concentration ($x$). The constants $C_{11}$ first rises, then slowly decreases with Te concentration. $C_{33}$ increases with $x$. $C_{12}$ and and $C_{66}$ both decrease, whereas $C_{13}$ and $C_{44}$ remain almost unchanged as Te doping level increases.

The FeSe$_{1-x}$Te$_x$ phases under consideration are usually synthesized in the form of polycrystalline substances. Thus the numerical estimates of the mechanical characteristics in these phases are desirable. The Voigt-Reuss-Hill procedure [24-26] is frequently used for estimating the elastic characteristics of polycrystalline materials using $C_{ij}$-values for single crystals. Hill [26] proved that the Viogt and Reuss equations represent upper and lower limits of true polycrystalline constants. He showed that the polycrystalline moduli are the arithmetic mean values of the moduli (monocrystalline values) in the Voigt ($B_V$, $G_V$) and Reuss ($B_R$, $G_R$) approximation and thus given by

$$G_H \equiv G = \tfrac{1}{2}(G_V + G_R), \qquad B_H \equiv B = \tfrac{1}{2}(B_V + B_R) \tag{1}$$

The Young's modulus $Y$ and Poisson's ratio $\nu$ are then computed from these values using the following relationship: $Y = 9BG/(3B + G)$, $\nu = (3B - Y)/6B$. The bulk moduli ($B_H$), shear moduli ($G_H$), Young's moduli ($Y$), and Poisson's ratios ($\nu$) in the Voigt-Reuss-Hill approximation for polycrystalline FeSe$_{1-x}$Te$_x$ are presented in Table 2. The expression for Reuss and Voigt moduli can be found elsewhere (see [27]). We compare the values with those of polycrystalline elastic moduli of LaFeAsO, LiFeAs and SrFe$_2$As$_2$ [28, 29]. The calculated values of SrFe$_2$As$_2$ reported in Table 5 of ref. [28] contain typographical error. The correct $G$ value (in Voigt approximation) has been estimated from where correct values of $B/G$, $Y$ and $\nu$ are reported in the table. The calculated moduli of 38.4 - 42.6 GPa reveal extremely FeSe$_{1-x}$Te$_x$ ($x = 0 – 1.0$) as soft materials. These values are smaller than the bulk moduli (122 - 210 GPa) of other well known superconducting materials such as MgB$_2$, MgCNi$_3$, YBCO and YNi$_2$B$_2$C [29], and compare well with the measured value of 30.7 GPa for α-FeSe obtained in complementary synchrotron X-ray diffraction at 16 K [7].

Table 2. Calculated elastic parameters for polycrystalline FeSe$_{1-x}$Te$_x$ in comparison to LaFeAsO, SrFe$_2$As$_2$, and LiFeAs ceramics [28, 29].

| Phase | $B_H = B$ (GPa) | $G_H = G$ (GPa) | $B/G$ | $Y$ (GPa) | $\nu$ | Ref |
|---|---|---|---|---|---|---|
| FeSe | 38.4 | 31.5 | 1.22 | 73.0 | 0.179 | This |
| FeSe$_{0.75}$Te$_{0.25}$ | 38.4 | 30.7 | 1.25 | 71.0 | 0.19 | This |
| FeSe$_{0.50}$Te$_{0.50}$ | 40.8 | 30.8 | 1.33 | 72.0 | 0.20 | This |
| FeSe$_{0.25}$Te$_{0.75}$ | 42.6 | 29.0 | 1.67 | 70.0 | 0.22 | This |
| FeTe | 40.7 | 29.8 | 1.37 | 70.0 | 0.21 | This |
| LaFeAsO | 97.9 | 56.2 | 1.74 | 141.5 | 0.259 | [28] |
| LiFeAs | 92.8 | 58.0 | 1.60 | 144.0 | 0.241 | [28] |
| SrFe$_2$As$_2$ | 61.7 | 32.1 | 1.92 | 82.1 | 0.28 | [28,29] |

According to Pugh [30], a given material is classified as brittle if $B/G < 1.75$. The results of our calculations give $B/G \sim 1.22 - 1.67$, which should imply that these materials occur below the boundary of brittle to plastic state. At the same time, $SrFe_2As_2$ (which is characterized by $B/G \sim 1.92$) must behave as a plastic material. The Poisson's ratios of covalent systems are known to be small ($\nu \sim 0.1$), while those of ionic crystals are $\nu \sim 0.25$. In addition, the covalent and ionic systems obey the relationships: $G \sim 1.1\ B$ and $G \sim 0.6\ B$, respectively. As can be seen from the data presented in Table 2, $FeSe_{1-x}Te_x$ should have mixed covalent and ionic bonds, whereas all of the LaFeAsO, LiFeAs, and $SrFe_2As_2$ ceramics belong to systems with predominantly ionic bonds.

The Debye temperature is proportional to the mean sound velocity $v_m$ [31]:

$$\Theta_D = \frac{h}{k} \sqrt[3]{\frac{3n}{4\pi V_0}}\ v_m \qquad (2)$$

where $h$ is Planck's constant, $k$ the Boltzmann's constant, $V_o$ the volume of unite cell and $n$ the number of atoms in unit cell, $v_m$ is the average wave velocity and can be obtained from the transverse $v_t$ and longitudinal wave velocity $v_l$, respectively (see [27]). We list the calculated $v_t$, $v_l$, $v_m$ and $\Theta_D$ in Table 3.

Table 3. The transver, longitudinal, average elastic wave velocities, and Debye temperature for parent and doped compounds.

| Compound | $v_t$ (km/s) | $v_l$ (km/s) | $v_m$ (km/s) | $\Theta_D$ (K) |
|---|---|---|---|---|
| [a]$Fe_{1.01}Se$ | - | - | 2.05 | 240 |
| [b]$FeSe$ | 2.49 | 3.58 | 2.50 | 285 |
| [b]$FeSe_{0.75}Te_{0.25}$ | 2.15 | 3.47 | 2.19 | 250 |
| [b]$FeSe_{0.5}Te_{0.5}$ | 2.11 | 3.45 | 2.16 | 246 |
| [b]$FeSe_{0.25}Te_{0.75}$ | 2.04 | 3.4 | 2.07 | 230 |
| [b]$FeTe$ | 2.00 | 3.30 | 2.05 | 233 |

[a] [32]; [b] This.

### 3.3 Electronic Properties

The calculated non-spin-polarized band structures, electronic densities of states (DOS) of $FeSe_{1-x}Te_x$ ($x = 0 - 1$) are given in Fig. 2 and Fig. 3, respectively. Fig. 2(a) shows the calculated band dispersions of iron-selenium (FeSe). The ten Fe-$d$ states are localized in an energy window extending from +2.9 eV to -2.6 eV around the Fermi level, where they give the dominant contribution to the DOS. The anion $p$ bands are seen to be located well below the Fermi level. As seen in the DOS in Fig. 3a these states are only hybridized modestly with the Fe-$d$ states. The general features are as observed in the work of Subedi *et al.* [8]. Further one also finds a strong qualitative similarity between FeSe and the FeAs-based superconductors [29, 33].

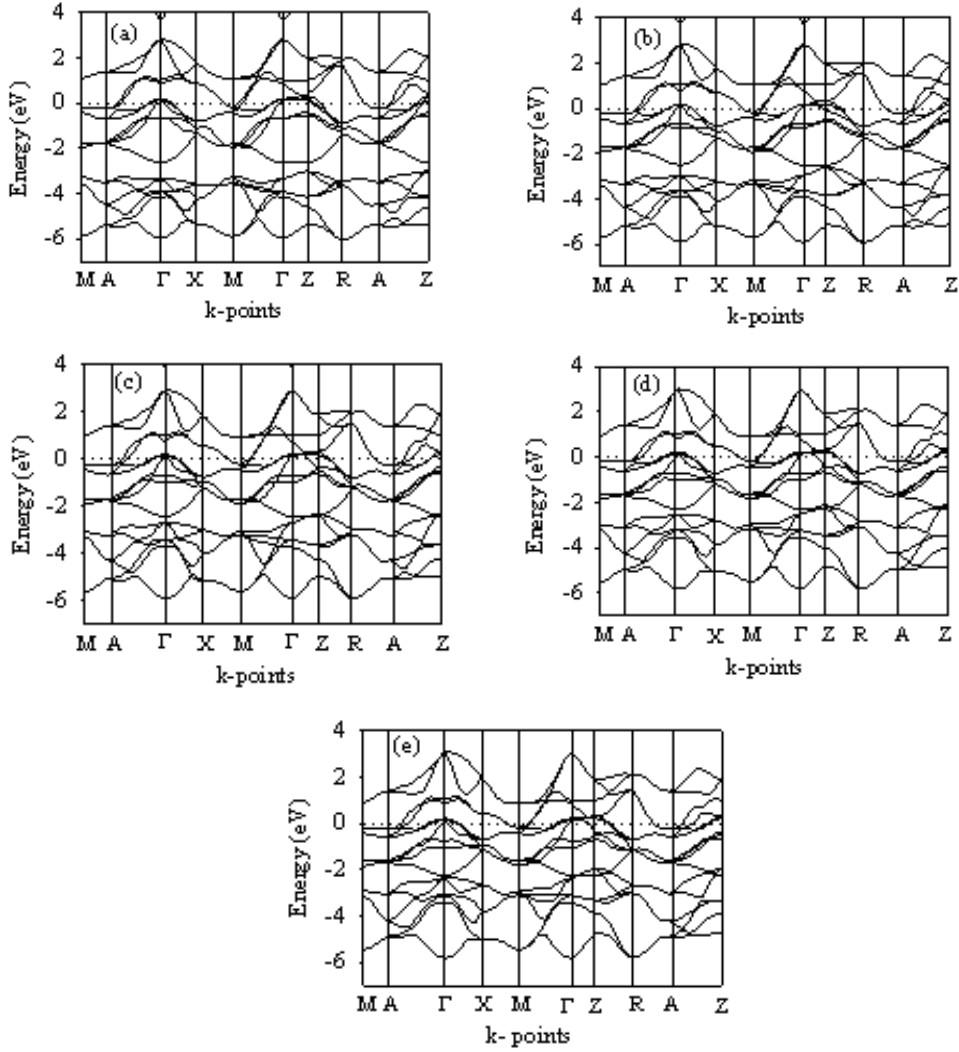

Fig.2. Electronic band structure of FeSe$_{1-x}$Te$_x$, for (a) $x = 0$, (b) $x = 0.25$, (c) $x = 0.50$, (d) $x = 0.75$, and (e) $x = 1.0$.

As indicated earlier, doping has an effect on the superconductivity of FeSe$_{1-x}$Te$_x$ ($x = 0 - 1$), whereby $T_c$ reaches a maximum of ~ 15 K at ~ 50-70% Te substitution, and then decreases again. In order to study the effects on electronic properties of substitution of Te on the Se sub-lattice we now analyze the band structures and density of states of FeSe$_{1-x}$Te$_x$ for various values of doping levels in Fig. 2 (b,c,d,) and Fig. 3 (b,c,d), respectively. With reference to band structure for $x = 0$ (Fig. 2a) the substitution of Te creates states in the gap region with substantial disorder in the Se-derived bands. It is expected from a consideration of atomic size that increased substitution of Te would lead to more disorder. The states created in the gap between Se- and Fe-derived bands can be clearly seen in Fig. 2 (b,c,d). As the doping level reaches ~0.5, there is a much more rearrangement of the bands near Fermi level. This continues to happen up to doping level ~0.75.

The total DOS increases progressively and the values change by 5.8% as the substitution level changes from $x = 0$ to $x = 1.0$ (Fig. 3). Further examination of the DOS shows that Fe $d$ manifold is split into two main peaks. These are separated by a pseudogap, with $E_F$ occurring towards the bottom. This pseudogap occurs at an electron count of six per Fe, corresponding to the $d$ electron count of $Fe^{2+}$. The relative importance of direct Fe-Fe interactions in the formation of the band structure is indicated by the position of pseudogap. One can notice the change of shape and position of the gap as the doping level is increased.

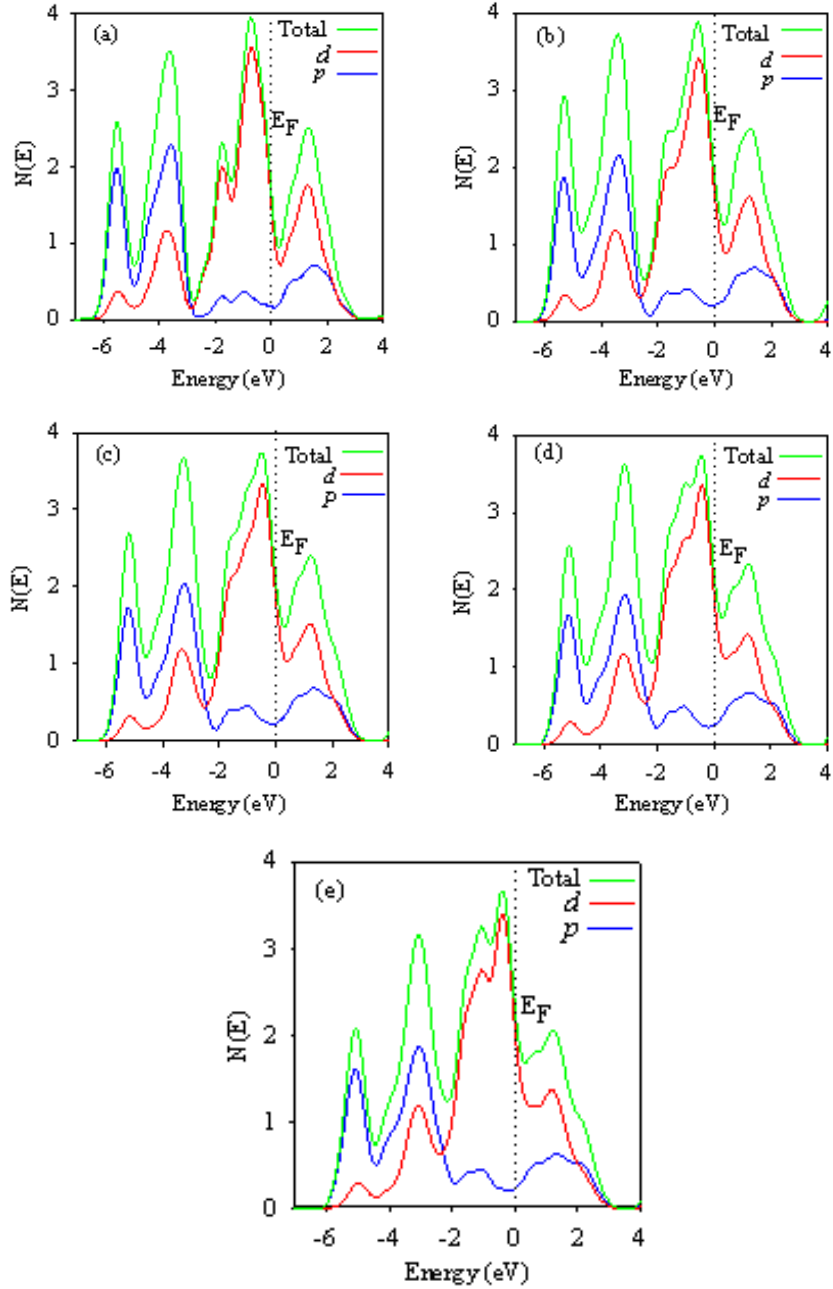

Fig. 3. Density of States of $FeSe_{1-x}Te_x$, for (a) $x = 0$, (b) $x = 0.25$, (d) $x = 0.75$, and (e) $x = 1$.

As $x$ increases to 1.0, we get the ordered alloy FeTe. In Fig. 2e and 3e, we show the band structure, and the density of states of FeTe alloy, which are similar to those of [8] except for the gap between Se- and Fe-derived bands around -2.5 eV below $E_F$ in FeSe. For FeSe we find that the gap between Se- and Fe-derived bands around 2.5 eV below $E_F$ in FeSe to be 0.48 eV (Fig. 2a). This may be compared with the values of 0.3, 0.4 eV (with enlarged muffin tin spheres) predicted by FPLMTO method [18]. Further in this case Te-$p$ bands hybridize with Fe-$d$ at around -2.2 eV along Γ- Z line while in FeSe ($x = 0$) Se-$3p$ is much separated from the Fe-$3d$ states.

We see different features along Γ-Z at energies above the Fermi level. Although there are no band crossings in the case of FeSe ($x$=0), there are significant band crossings for FeTe ($x = 1$) at about + 0.3 eV. Another notable difference exists at around the X point above the Fermi level where the parabolic band along X-M-Γ is flattening in FeTe. It is also noted that unlike FeSe there are no band crossings across the X point in the range of 0.48 eV to1.94 eV. These differences above the Fermi level indicate a different Fermi surface topology induced by Te or electron doping.

The rearrangement of bands in FeTe is different and thus is expected to have changes in the shape and nesting of the Feri surface (FS). In fact it has been shown by Singh [18] that FeTe ($x = 1.0$) reveal rather smaller FS nesting at the Γ-X-M plane compared with FeSe.

It seems clear from theoretical calculations and experimental characterizations that the superconductivity in Fe-pnictides is not mediated by phonons (see [8]). Therefore, attempts are being directed to understand the superconductivity in Fe-pnictides in terms of spin fluctuations and related theories. In spin-fluctuation theories [34], in addition to the proximity of the system to a magnetic instability, the FS and, in particular, the FS nesting plays a crucial role in enhancing the magnetic interactions. Nesting is a property which originates from the shape of the FS, and it enhances quasiparticle scattering along particular directions in momentum space. As mentioned above the FS of the system can be changed by chemical substitutions, or even by electron or hole doping. The changes in the FS topology are the main driving force behind the observed trends in band structure. The substitution of Te enhances the possibility of Fermi surface (FS) nesting, especially in FeSe$_{0.5}$Te$_{0.5}$, despite the disordering the Se-derived bands [18]. In order to determine whether a given FS supports nesting, an intimate knowledge of the full three-dimensional morphology is required. Instead of doing such detailed investigation we would use results of a detailed theoretical study by Singh [18]. The observed band structure for $x = 0.5$ induced by the changes in the FS topology lead to the effect such as maximizing the nesting of Fermi surface in the Γ-X-M plane [18]. We note that at this Te concentration the observed superconducting $T_c$ is maximum. The correlation between favourable nesting and superconducivity observed in BaFe$_2$(As$_{1-x}$P$_x$)$_2$ provides evidence for the importance of nesting in understanding Fe-pnictide superconductivity, as suggested by several researchers [35]. This may provide a possible basis for understanding why $T_c$ in FeSe$_{1-x}$Te$_x$ is found to be maximum in the doped level ~ 0.5 as has been discussed in this section.

## 4. Conclusion

In the present work the effect of doping level on elastic, electronic and other related properties in FeSe$_{1-x}$Te$_x$, including for $x$ value for which $T_c$ is maximum, has been investigated theoretically based on the first-principles DFT total energy calculations. The elastic properties for mono- and poly-crystalline FeSe$_{1-x}$Te$_x$ system are predicted for the first time. Our analysis showed that the system is stable. All the alloys are soft materials ($B < 45$ GPa) with high compressibility and exhibit a somewhat brittle behaviour. The higher concentration of Te is found to induce weaker coupling between each neighboring plane, resulting in more layer-like characteristics.

The band structure for the ordered alloy FeSe has a strong qualitative similarity with the FeAs-based superconductors. The alloys with $x = 0$ and $x = 1$ have different features along $\Gamma$-$Z$ and also across the X point in the band structure at energies above the Fermi level. These differences above the Fermi level indicate a different Fermi surface topology induced by electron doping. The characteristics of crossing Fermi level, rearrangement, and creation of states in the gap region in the Se-derived bands due to doping ($0.5 < x < 0.75$) indicate a more positive effect such as maximizing the nesting of Fermi surface in the $\Gamma$-X-M plane. A correlation between favourable nesting and superconducivity found in $BaF_2(As_{1-x}P_x)_2$ may also be operative in $FeSe_{1-x}Te_x$, which may provide an explanation of why $T_c$ is maximum for Te concentration of ~ 0.5.